\documentclass[prl,preprint, superscriptaddress, onecolumn]{revtex4}

\usepackage[utf8]{inputenc}
\usepackage{hyperref}
\usepackage{graphicx}
\usepackage{xcolor}
\usepackage{natbib}
\usepackage{amsfonts, amsmath, bm, graphicx}
\usepackage{siunitx}
\usepackage{float}
\usepackage{color,soul}
\usepackage{textgreek}
\usepackage{textcomp}
\usepackage{gensymb}

\DeclareSIUnit\angstrom{\text {Å}}
\let\DeclareUSUnit\DeclareSIUnit

\DeclareUSUnit\dBm{dBm}

\begin{document}
\author{Atul Pandey}
\affiliation{Max Planck Institute of Microstructure Physics, Weinberg 2, 06120 Halle, Germany}
\affiliation{Institute of Physics, Martin Luther University Halle-Wittenberg, Von-Danckelmann-Platz 3, 06120 Halle, Germany} 
\author{Prajwal Rigvedi}
\affiliation{Max Planck Institute of Microstructure Physics, Weinberg 2, 06120 Halle, Germany}
\author{Edouard Lesne}
\affiliation{Max-Planck-Institute for Chemical Physics of Solids, Nöthnitzer Straße 40, 01187 Dresden, Germany} 
\author{Jitul Deka}
\affiliation{Max Planck Institute of Microstructure Physics, Weinberg 2, 06120 Halle, Germany}
\author{Jiho Yoon}
\affiliation{Max Planck Institute of Microstructure Physics, Weinberg 2, 06120 Halle, Germany}
\author{Wolfgang Hoppe}
\affiliation{Institute of Physics, Martin Luther University Halle-Wittenberg, Von-Danckelmann-Platz 3, 06120 Halle, Germany}
\author{Chris Koerner}
\affiliation{Institute of Physics, Martin Luther University Halle-Wittenberg, Von-Danckelmann-Platz 3, 06120 Halle, Germany} 
\author{Banabir Pal}
\affiliation{Max Planck Institute of Microstructure Physics, Weinberg 2, 06120 Halle, Germany}
\author{James M. Taylor}
\affiliation{Institute of Physics, Martin Luther University Halle-Wittenberg, Von-Danckelmann-Platz 3, 06120 Halle, Germany}
\author{Stuart S. P. Parkin}
\affiliation{Max Planck Institute of Microstructure Physics, Weinberg 2, 06120 Halle, Germany}
\author{Georg Woltersdorf}
\affiliation{Institute of Physics, Martin Luther University Halle-Wittenberg, Von-Danckelmann-Platz 3, 06120 Halle, Germany} 
\affiliation{Max Planck Institute of Microstructure Physics, Weinberg 2, 06120 Halle, Germany}
\email{georg.woltersdorf@physik.uni-halle.de}

\date{\today}

\title{
Switching of magnetic domains in a noncollinear antiferromagnet at the nanoscale}

\begin{abstract}

 Antiferromagnets that display very small stray magnetic field are ideal for spintronic applications. Of particular interest are non-collinear, chiral antiferromagnets of the type $\mathrm{Mn_3X}$ (X=Sn, Ge), which display a large magnetotransport response that is correlated with their antiferromagnetic ordering.  The ability to read out and manipulate this ordering is crucial for their integration into spintronic devices. These materials exhibit a tiny unbalanced magnetic moment such that a large external magnetic field can, in principle, be used to set the material into a single antiferromagnetic domain. However, in thin films of Mn\textsubscript{3}Sn, we find that such fields induce only a partial magnetic ordering. By detecting two orthogonal in-plane components of the magnetic order vector, we find that the non-switchable fraction has a unidirectional anisotropy. This also enables us to visualize switching along multiple easy axes in $\mathrm{Mn_3Sn}$. Studying the switching at the nanoscale allows us to correlate the pining behavior to crystal grain boundaries in the $\mathrm{Mn_3Sn}$ nanowire structures. 
\end{abstract}

\maketitle

Antiferromagnetic materials are promising candidates for high performance spintronic devices due to their fast dynamics, stability against external perturbations, and ability to miniaturize components due to the lack of stray fields\cite{AS_AIP_Gomonay2014,AS_MacDonald2011,AS_RevModPhys_2018,AS_Nat_NanoTech_Jungwirth2016}. Of particular interest are antiferromagnetic materials with topologically-protected bandstructures\cite{TAS_Nat_Phy_mejkal2018,TAS_Bonbien2021}. These include noncollinear antiferromagnets (NCAFs). Here, triangular spin textures with an inherent chirality\cite{TST_Brown1990,TST_npj_Park2018} lead to the emergence of Weyl points in their bandstructure\cite{Weyl_ferm_nat_matKuroda2017,Weyl_mejkal2017}. These, in turn, act as a strong internal effective magnetic field\cite{berry_cur_field_Kbler2014,berry_cur_field_PRL_Chen2014}. The Berry curvature associated with such Weyl points leads NCAFs to exhibit magneto-transport and magneto-optical effects analogous to conventional ferromagnetic materials. These include the anomalous Hall effect\cite{NCAFM_AFE_Nat_Nakatsuji2015}, anomalous Nernst effect (ANE)\cite{NCAFM_ANE_Nat_phy_Ikhlas2017}, magneto-optical Kerr effect (MOKE)\cite{NCAFM_MOKE_nat_photo_Higo2018,NCAFM_MOKE_Uchimura2022,NCAFM_MOKE_PhysRevB.92.144426}and x-ray magnetic circular dichroism\cite{NCAFM_XMCD_PhysRevB.104.134431,NCAFM_XMCD_Nat_com_Kimata2021}.

The behavior of the magnetic order in response to external stimuli, for example, magnetic fields or spin torques, can be understood in terms of a magnetic octupole order parameter (MOOP)\cite{NCAFM_XMCD_Nat_com_Kimata2021,MOOP_Chen2023,MOOP_Nakatsuji2022}. This is defined by a vector given by the sense of rotation of the chiral spin texture within a kagome plane. The advantage of NCAFs is that magnetic octupole-driven electrical or optical read-out signals are much larger than those obtained from collinear antiferromagnetic materials\cite{`moop_resp_Chen2021,NCAFM_AFE_Nat_Nakatsuji2015,NCAFM_ANE_Nat_phy_Ikhlas2017}. Being able to switch the orientation of the MOOP in NCAFs and read its state electrically or optically brings such devices closer to realization. However, to properly understand the switching mechanisms at play, it is important to study the dynamics and magnetic domain structures in NCAFs. Specifically, it is required to image their antiferromagnetic domains, which correspond to regions of triangular spin texture with the same chiral sense of rotation but oriented along one of three symmetry related directions in the hexagonal plane. Because these domains are typically very small\cite{dome_size_Cheong2020,NCAFM_MOKE_nat_photo_Higo2018,dom_size_Krizek2022}, imaging must be performed with nanometer resolution. 
Previously, scanning ANE (SANE) measurements have been performed to observe magnetic field-induced switching of the MOOP vector in NCAFs at micrometer scales\cite{SANE_Reichlova2019,SANE_Johnson2022}. This is achieved by using a focused laser beam to create a localized temperature gradient $\mathbf{\nabla} T$.This gradient generates an electrical ANE response ($\mathbf{E}_{\mathrm{ANE}}$) transverse to both $\mathbf{\nabla} T$ and the MOOP vector (\textbf{M}).

In this letter, we perform spatially resolved ANE measurements to study the spatial distribution of the MOOP in a NCAF thin film of $\mathrm{Mn_3Sn}$. By detecting two orthogonal components of $\mathbf{E}_{\mathrm{ANE}}$, we visualize switching of the MOOP into arbitrary in-plane directions of $\mathrm{Mn_3Sn}$ on a sub-micron length scale. Since the spatial resolution is not sufficient to fully understand the inhomogeneous switching behavior we have extended the method, to map the magnetic domains with nanoscale spatial resolution. We achieve this by focusing a laser beam on to the apex of an atomic force microscope (AFM) tip to create a heat gradient on the nanometer scale, allowing us to correlate the inhomogenous switching behavior of the NCAF to it's grain structure.

\begin{figure}[H]
\centering
\includegraphics[width=1\textwidth]{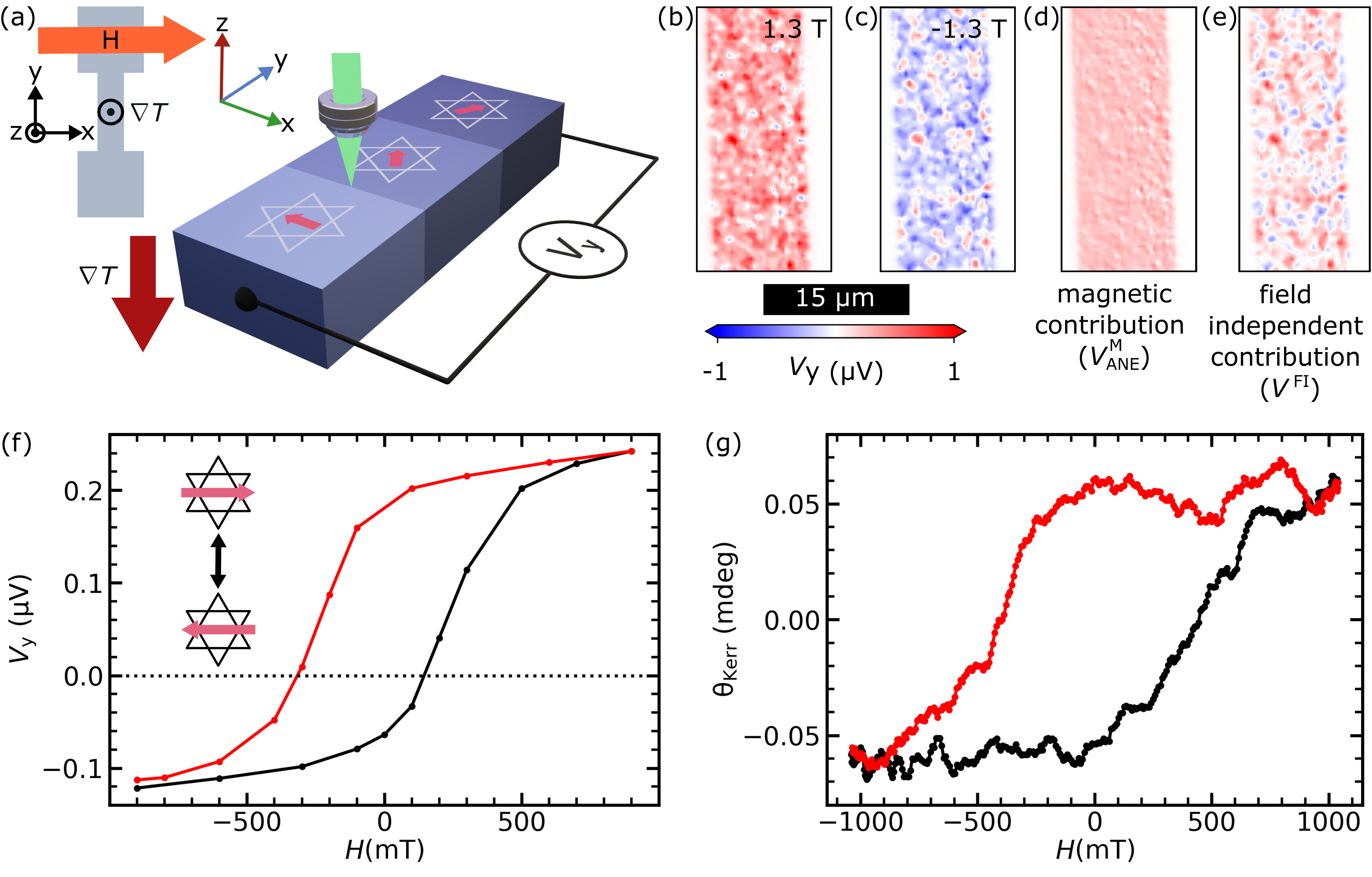}
\caption{(a) Schematic showing the device structure in the upper left and an illustration of the ANE microscope  on the right. The laser beam with a power of \SI{1}{mW} and a wavelength of \SI{532}{nm} is focused on to the sample using an objective lens with a numerical aperture of 0.7 to create an OOP temperature gradient (vertical red arrow). This generates an ANE voltage transverse to the temperature gradient and the MOOP vector (indicated by arrows on the slab). (b,c) ANE microscope image of a $\mathrm{Mn_3Sn}$ wire structure that has been field cooled in $\pm$1.3 T, respectively. (d) $V_{\mathrm{y}}$ in (c) is subtracted pixel by pixel from that in (b) and divided by 2 to obtain the map of $V^{\mathrm{M}}_{\mathrm{ANE}}$. (e) Map of $V^{\mathrm{FI}}$ obtained by taking pixel by pixel average of (b) and (c). (f) ANE-based hysteresis loop measured at \SI{125}{\degree C}. The inset schematically indicates the effect of the magnetic field on the magnetic order parameter. (g) Hysteresis at \SI{125}{\degree} obtained by longitudinal MOKE.} 
\label{fig:ff_method}
\end{figure}

We investigate a c-axis-oriented \SI{60}{nm} thick $\mathrm{Mn_3Sn}$ film grown using magnetron sputtering on a MgO (111) substrate (see supplementary
for methods). The film is patterned into a wire of width, $w=\SI{10}{\micro m}$, as shown schematically in \textbf{Fig. \ref{fig:ff_method}(a)}. The wire structure is illuminated with a laser beam to create an out-of-plane (OOP) temperature gradient ($\nabla T_{\mathrm{z}}$). This results in a ANE induced electrical response given by:
\begin{equation}\label{eq1}
    \mathbf{E}_{\mathrm{ANE}} = \mu_{\mathrm{0}}\cdot S_{\mathrm{ANE}}\cdot \mathrm{\mathbf{M}} \times \mathbf{\nabla}T = \mu_{\mathrm{0}}\cdot S_{\mathrm{ANE}}\cdot M_{\mathrm{\mathrm{y}}}\cdot \nabla T_{\mathrm{z}}\cdot\mathbf{\hat{x}}+\mu_{\mathrm{0}}\cdot S_{\mathrm{ANE}}\cdot M_{\mathrm{\mathrm{x}}}\cdot \nabla T_{\mathrm{z}}\mathbf{\cdot\hat{y}}
\end{equation}
where $S_{\mathrm{ANE}}$ is the ANE coefficient and \textbf{M} is the in-plane orientation of the MOOP unit vector and $\mu_0$ is the permeability constant. First, we measure the y-component of $\mathbf{E}_{\mathrm{ANE}}$ ($E_{\mathrm{ANE}}^{\mathrm{y}}=\mu_{\mathrm{0}}\cdot S_{\mathrm{ANE}}\cdot M_{\mathrm{\mathrm{x}}}\cdot \nabla T_{\mathrm{z}}$) by detecting the ANE-induced voltage in a wire oriented along the y-direction ($V_{\mathrm{y}} \propto E_{\mathrm{ANE}}^{\mathrm{y}}$). The spatially resolved measurement of $V_{\mathrm{y}}$ allows us to create a map of the x-component of the order parameter ($M_{\mathrm{x}}$). 

First, we consider the case where the laser beam is focused using a microscope objective. We switch the MOOP vector from $+\mathbf{\hat{x}}$ to $-\mathbf{\hat{x}}$, as shown schematically in the inset of \textbf{Fig. \ref{fig:ff_method}(f)}, by applying a magnetic field along the positive or negative x-directions at an elevated temperature (\SI{125}{\degree C}). At this temperature the coercivity is reduced, so that a \SI{\pm 1.3}{T} magnetic field is sufficient to saturate the MOOP. The effect of temperature and the applied magnetic field on the switching behavior is shown in supplementary \textbf{Fig. S2}. The $\mathrm{Mn_3Sn}$ device is then cooled to room temperature in the presence of the magnetic field, where spatially resolved ANE measurements are performed (thus avoiding temperature-induced drift). The spatial distribution of $V_{\mathrm{y}}$ is shown in \textbf{Fig. \ref{fig:ff_method}(b,c)}. When the device is cooled in the presence of \SI{1.3}{T} or \SI{-1.3}{T}, the majority of regions show positive or negative signs of the measured voltage, respectively. This sign change corresponds to the reversal of the ANE signal because of the field-induced switching of the MOOP vector. 

There are small areas where the signal polarity is insensitive to the field. We attribute these to domain pinning or grain boundaries. Domain pinning results in a field-independent ANE voltage ($V_{\mathrm{ANE}}^{\mathrm{FI}}$), while grain boundaries result in a non-magnetic Seebeck-effect (SE) voltage ($V_{\mathrm{SE}}^{\mathrm{NM}}$)\cite{SNOM_otani_PhysRevLett.132.216702}. The magnetic field-dependent ANE signal ($V^{\mathrm{M}}_{\mathrm{ANE}}$) can be separated by subtracting two images where the MOOP is field polarized in opposite directions  (\textbf{Fig. \ref{fig:ff_method}(d)}). Similarly, by adding these images we obtain a spatial map of field-independent contributions. The resulting map is shown in \textbf{Fig. \ref{fig:ff_method}(e)}, and reveals that there is a significant field-independent contribution ($V^{\mathrm{FI}} =V_{\mathrm{ANE}}^{\mathrm{FI}}+V_{\mathrm{SE}}^{\mathrm{NM}}$) that is present throughout the wire structure and is bipolar.

The magnetic field dependence of the MOOP reversal is further studied by hysteresis loop from the ANE. \textbf{Fig. \ref{fig:ff_method}(f)} shows the average $V_{\mathrm{y}}$ measured over a \SI{10}{\micro m}$\times$\SI{20}{\micro m} area as a function of the applied magnetic field at a temperature of \SI{125}{\degree C}. This demonstrates the field-induced switching of the magnetic order and is validated by comparing it with the hysteresis loop obtained using longitudinal MOKE, as shown in \textbf{Fig. \ref{fig:ff_method}(g)}.%
 
We observe an offset in $V_{\mathrm{y}}$ in the ANE-based hysteresis loop. This results from the field-independent contributions, either through the non-magnetic Seebeck effect $V_{\mathrm{SE}}^{\mathrm{NM}}$, or as a result of partial field-induced switching, $V_{\mathrm{ANE}}^{\mathrm{FI}}$. In the latter case, an offset will arise if the population of pinned domains with MOOP vector along $+\mathbf{\hat{x}}$ is higher relative to that along $-\mathbf{\hat{x}}$. 

\begin{figure}[H]
\centering
\includegraphics[width=1\textwidth]{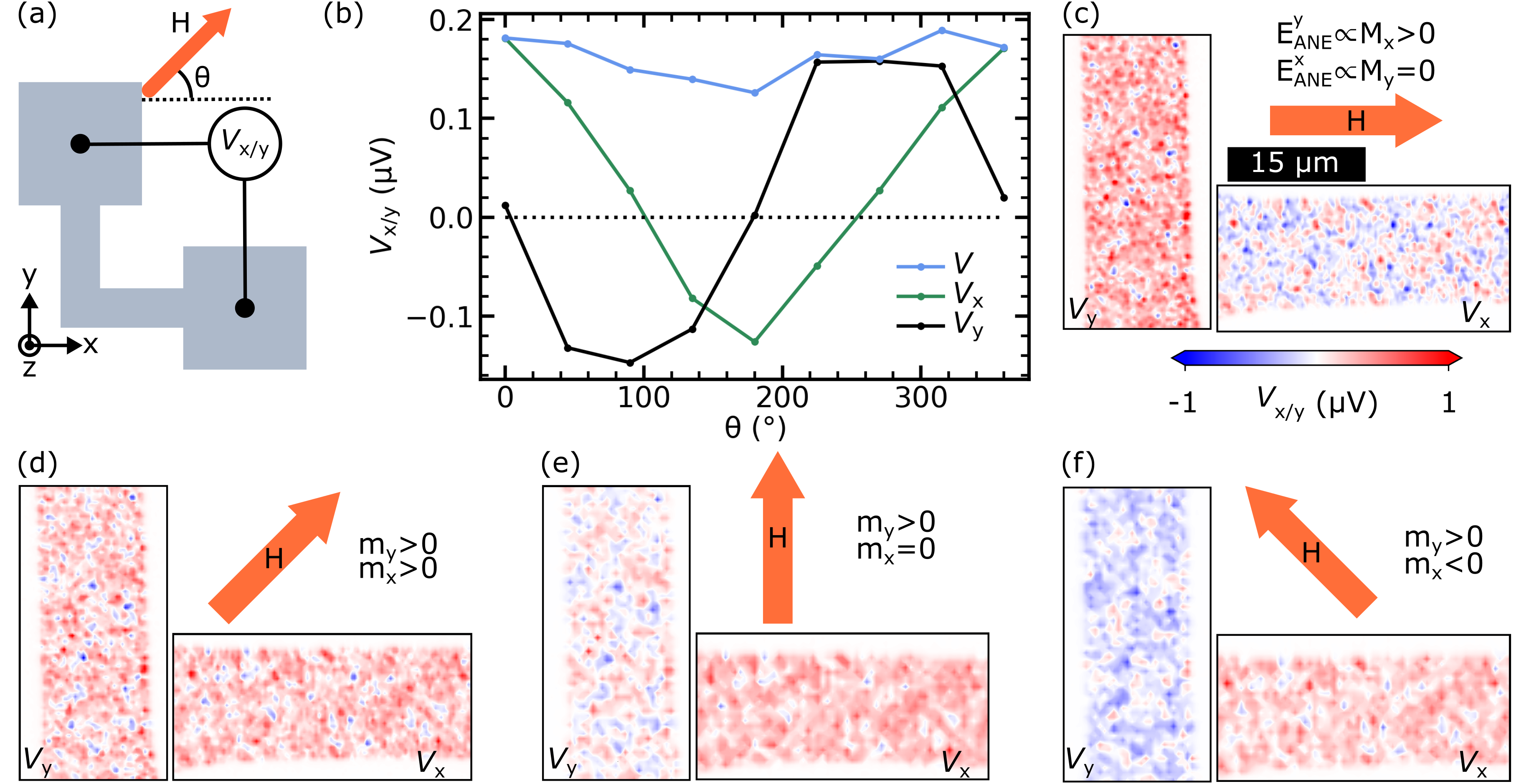}
\caption{(a) Schematic of the device for angle-dependent field measurements. (b) Average voltage signal measured in the horizontal ($V_{\mathrm{x}}$) and vertical arms ($V_{\mathrm{y}}$) of the device against the in-plane field angle $\theta$. $\theta = \SI{90}{\degree}$ corresponds to the positive y-direction whilst $\theta = \SI{0}{\degree}$ corresponds to the positive x-direction. (c-f) ANE microscopy domain images for different in-plane field angles, as indicated by the arrows: (c) $\theta = \SI{0}{\degree}$, (d) $\theta = \SI{45}{\degree}$, (e) $\theta = \SI{90}{\degree}$, (f) $\theta = \SI{135}{\degree}$. The left panels show scans of a \SI{15}{\micro m}$\times$\SI{30}{\micro m} area in the vertical arm, whilst the right panels show scans of a \SI{30}{\micro m}$\times$\SI{15}{\micro m} area in the horizontal arm.} 
\label{fig:ff_six_dom}
\end{figure}

To verify the presence of partial field-induced switching, we study the angular dependence of the measured signal. For this set of measurements, we fabricate a device with two orthogonal arms as shown schematically in \textbf{Fig. \ref{fig:ff_six_dom}(a)}. The voltage measured in the horizontal ($V_{\mathrm{x}}$), and vertical ($V_{\mathrm{y}}$) arms of this device allows us to deduce $M_{\mathrm{y}}$ and $M_{\mathrm{x}}$ ($\mathbf{M}=M_{\mathrm{x}}\mathbf{\hat{x}}+M_{\mathrm{y}}\mathbf{\hat{y}}$), respectively (see equation\eqref{eq1}). Magnetic order is changed by applying a \SI{1.3}{T} in-plane magnetic field along different angles, $\theta$, at \SI{125}{\degree C}. Subsequently, we image the magnetic domains after field cooling to room temperature. Average values of $V_{\mathrm{x}}$ ($V_{\mathrm{y}}$), which are directly proportional to the two components of MOOP $M_{\mathrm{y}}$ ($M_{\mathrm{x}}$), are plotted against $\theta$ in \textbf{Fig. \ref{fig:ff_six_dom}(b)}. A \SI{90}{\degree} phase shift between $V_{\mathrm{x}}$ and $V_{\mathrm{y}}$ confirms that they are related to the two orthogonal component of \textbf{M}. Both show periodic behavior with $\theta$, indicating that the MOOP vector switches along multiple directions.

We investigate this further using spatially resolved measurements in the two arms to visualize the magnetic octupole switching in different directions. When the magnetic field is applied along the positive x-direction, (\textbf{Fig. \ref{fig:ff_six_dom}(c)}), the MOOP orients with the field, resulting in a positive y-component of $\mathbf{E}_{\mathrm{ANE}}$. This is indicated by the red color in the visual representation of the vertical arm: the majority of regions appear as red due to the dominance of the field-dependent ANE signal resulting from positive $M_{\mathrm{x}}$. On the other hand, the net $M_\mathrm{y}$, and therefore x-component of $\mathbf{E}_{\mathrm{ANE}}$, is zero. Consequently, the signal in the horizontal arm shows only a non-magnetic signal $V^{\mathrm{FI}}$. When the field is applied at \SI{45}{\degree} with respect to the x-axis, we observe a dominant red contrast in both arms (\textbf{Fig. \ref{fig:ff_six_dom}(d)}). This indicates that both $M_\mathrm{x}$ and $M_\mathrm{y}$ are positive, due to the MOOP following the rotated field. Application of the field along the y direction results in positive $M_\mathrm{y}$ and net zero $M_\mathrm{x}$ (\textbf{Fig. \ref{fig:ff_six_dom}(e)}), whilst applying the field at \SI{45}{\degree} from the y-axis results in positive $M_\mathrm{y}$ and negative $M_\mathrm{x} (\textbf{Fig. \ref{fig:ff_six_dom}(f)})$.

These images demonstrate that the MOOP can be aligned along multiple in-plane directions by a magnetic field.
As shown by X-ray diffraction pole scans of our epitaxial film (supplementary \textbf{Fig. 1}), our $\mathrm{Mn_3Sn}$ samples are indeed hexagonal and well oriented. Thus, one can expect them to show three magnetic easy axes along the $<\Bar{2}\Bar{1}10>$ directions \cite{TST_Brown1990,six_dom_Pal2022,six_Sugimoto2020}. Applying a magnetic field in the plane of the film should therefore orient the MOOP along the closest of the possible six easy directions. 

However, if this field would achieve complete switching, the magnitude of \textbf{M} ($M=\sqrt{M_{\mathrm{x}}^2+M_{\mathrm{y}}^2}\propto V= \sqrt{V_{\mathrm{x}}^2+V_{\mathrm{y}}^2}$) would be independent of $\theta$. Instead, we observe a minimum in $V$, and therefore in the magnitude of \textbf{M}, at $\theta=$\SI{180}{\degree} in (\textbf{Fig. \ref{fig:ff_six_dom}(b)}). This indicates that the MOOP only partially aligns with the magnetic field and that this alignment is most favorable along the positive x-direction. 

In these images we also observe the presence of domain structures smaller than the optical resolution of \SI{\approx700}{nm} of the ANE microscope \cite{arxiv}. To properly understand this switching process, we need to image magnetic domains at the nanoscale. This is achieved by applying a laser beam to the apex of a vibrating AFM tip. Here, the tip-sample interaction results in an enhanced optical near-field underneath the tip, creating a temperature gradient with a lateral spread on the nanometer length scale. The resulting ANE voltage is demodulated at the second harmonic of the tip frequency, which corresponds purely to the voltage generated by the near-field temperature gradient. We have previously demonstrated that this technique can image magnetic textures in ferromagnets with a spatial resolution of \SI{80}{nm} \cite{arxiv}. Here, we apply the same technique to image magnetic domains in a c-axis-oriented \SI{60}{nm} thick $\mathrm{Mn_3Sn}$ film patterned into a \SI{300}{nm}$\times$ \SI{10}{\micro m} nanowire. 

\begin{figure}[H]
\centering
\includegraphics[width=1\textwidth]{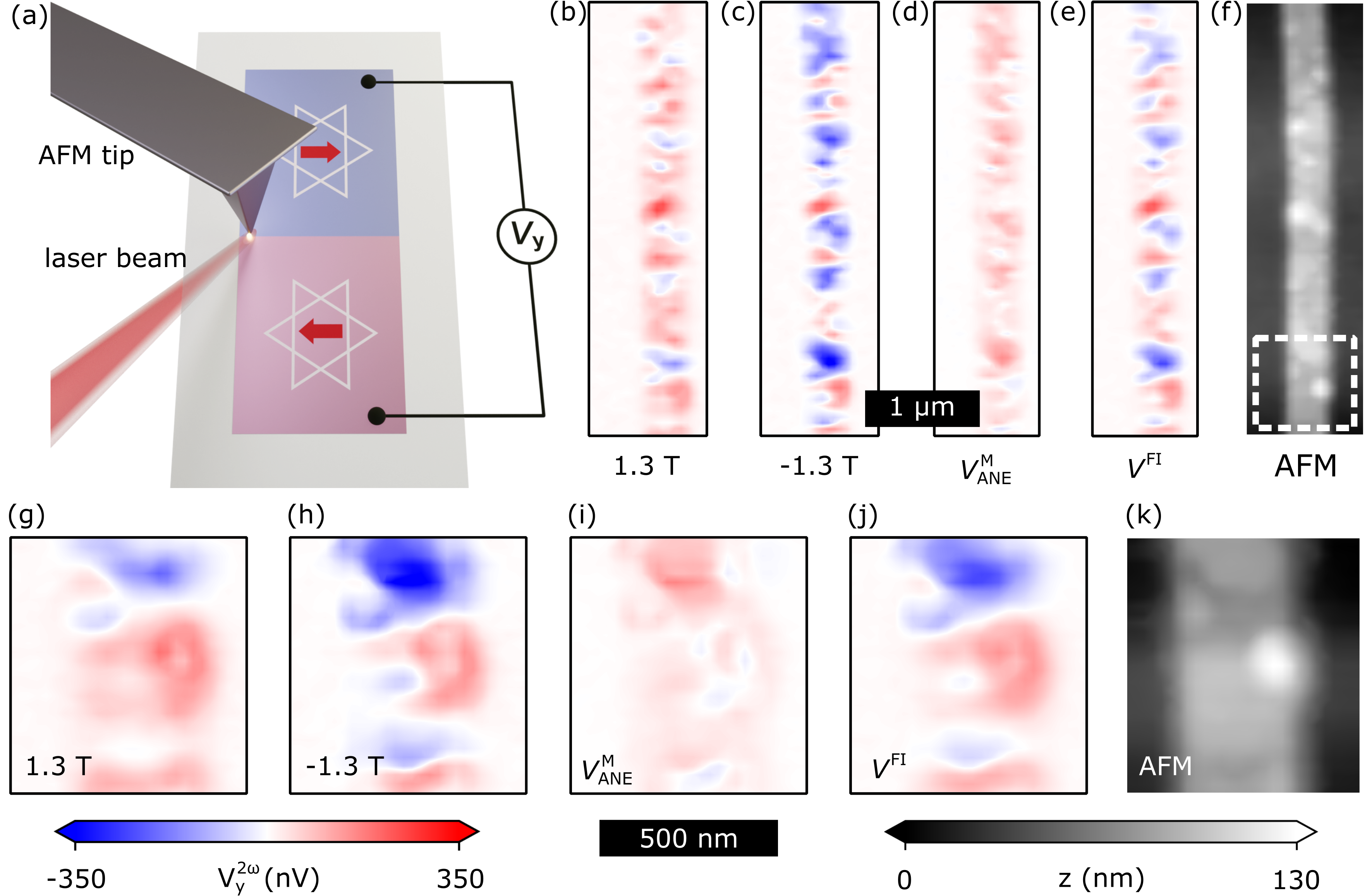}
\caption{(a) Schematic illustration of the near-field ANE microscope. (b-e) Scanning measurements performed at room temperature in a $\mathrm{Mn_3Sn}$ nanowire. The step size is \SI{50}{nm}. (b,c) ANE images after \SI{\pm1.3}{T} field cooling, respectively. (c) Magnetic part of near-field measured voltage $V_{\mathrm{ANE}}^{\mathrm{2\omega}}$. (d) Field-independent part of $V_{\mathrm{ANE}}^{\mathrm{2\omega}}$. (f)  AFM topography measured simultaneously with the ANE measurements in (c). (g-k) Scanning measurements performed in a smaller region, shown by the dotted square in (f), with a smaller \SI{20}{nm} step size. The images correspond to the same experimental conditions as (b-f).} 
\label{fig:nf_dom} 
\end{figure}

By following the same measurement procedure used in \textbf{Fig. \ref{fig:ff_method}(b-e)}, we image the  $\mathrm{Mn_3Sn}$ nanowires after positive (\textbf{Fig. \ref{fig:nf_dom}(b)})and negative (\textbf{Fig. \ref{fig:nf_dom}(c)}) field cooling. Subtracting and adding these images yields the magnetic (\textbf{Fig. \ref{fig:nf_dom}(d)}) and field-independent parts \textbf{Fig. \ref{fig:nf_dom}(e)} of $V_{\mathrm{y}}$, respectively. The simultaneously measured AFM topography, \textbf{Fig. \ref{fig:nf_dom}(f)}, provides insight into the structure and morphology of the thin film device.

We zoom in on a region containing multiple crystal grains, shown by the dotted square in \textbf{Fig. \ref{fig:nf_dom}(f)}, by measuring an ANE domain map with a smaller step size (higher pixel density) that gives better spatial resolution. The antiferromagnetic domain distribution after positive (\textbf{Fig. \ref{fig:nf_dom}(g)}) and negative (\textbf{Fig. \ref{fig:nf_dom}(h)}) field cooling is now imaged with a resolution of better than \SI{100}{nm}. 

At this length scale, \textbf{Fig. \ref{fig:nf_dom}(i,d)} shows that the magnetic contribution $V^{\mathrm{M}}_{\mathrm{ANE}}$ is spatially non-uniform. This is caused by chiral domains with a sense of rotation, some of which are pinned and can not be switched by the applied field, supporting our previous observation of pinned domains affecting the MOOP switching process in $\mathrm{Mn_3Sn}$. These pinned domains are, in turn, shown as regions with a higher magnitude of $V_{\mathrm{ANE}}^{\mathrm{FI}}$ in \textbf{Fig. \ref{fig:nf_dom}(j)}. By comparing these images with the topographic map of the nanowire in \textbf{Fig. \ref{fig:nf_dom}}(k), we observe that the regions with large $V^{FI}$ are correlated with positions of structural grains. We conclude, therefore, that the domains are pinned in place by crystalline grain boundaries in the thin film. 

Our observations are consistent with the switching behavior of collinear antiferromagnets observed using high resolution photoemission electron microscopy. Here, current-induced spin-orbit-torque in CuMnAs\cite{PEEM_SOT_CuMnAs_PhysRevLett.118.057701} and $\mathrm{Mn_2Au}$\cite{PEEM_SOT_Mn2AuPhysRevB.99.140409}, as well as applied magnetic fields of up to \SI{70}{T} in $\mathrm{Mn_2Au}$\cite{PEEM_field_Mn2AuPhysRevB.97.134429}, only induce partial switching. 

Another recent study of magnetic domain images in $\mathrm{Mn_3Sn}$ nanowires by near-field based ANE microscopy\cite{SNOM_otani_PhysRevLett.132.216702} also demonstrates magnetic field-dependent and independent contributions to the measured voltage signal. The field-independent contribution was attributed to a Seebeck effect-induced voltage at the grain boundaries. Our analysis of domain images together with AFM-based topography is consistent with this prior work. Furthermore, some regions were observed to show zero magnetic contribution ($V^{\mathrm{M}}_{\mathrm{ANE}}=0$) and these regions were ascribed to an OOP polarization of the MOOP. However, this is inconsistent with our recent work on ANE imaging of OOP magnetized nanowires \cite{arxiv}, where we show that the OOP magnetisation results in zero ANE signal only if the magnetization underneath the heat gradient spot is uniform. Therefore a non-uniform magnetization under the spot, for example, a domain wall or the edges of the nanowire, will result in a non-zero ANE signal due to the magnetization gradient\cite{ANE_PMA_edge_AIP_adv_2018,arxiv}. For these reasons, we ascribe the regions without field dependent signals $V^{\mathrm{M}}_{\mathrm{ANE}}$ in $\mathrm{Mn_3Sn}$ to be a result of pinned domains. 

In conclusion, we have imaged chiral antiferromagnetic domains in $\mathrm{Mn_3Sn}$ nanowires using near-field based scanning ANE microscopy with nanoscale spatial resolution. By observing two orthogonal components of $\mathbf{E}_{\mathrm{ANE}}$, we distinguish different orientations of the MOOP vector. Applying an in-plane magnetic field at various angles switches the MOOP vector along different directions that corresponds to the magnetic easy axis of $\mathrm{Mn_3Sn}$. Our nanoscale images reveal that only partial field-induced switching occurs, indicating domain pining in regions corresponding to crystal grains. The large crystal grains are a consequence of the high temperature annealing process that is needed for chemical ordering of the $\mathrm{Mn_3Sn}$ compound. Based on our results, we believe that smaller crystallites and more mobile domains may be achieved using alternative methods such as templated growth.\cite{Templated_growth_Filippou2018}

{\bf Acknowledgements}

We acknowledge financial support from the German research foundation (DFG) through collaborative research center (CRC) 227 (project B10), under Project-ID 328545488. 

{\bf Author contributions}
The contribution from the authors is as follows: 

Conceptualization: GW, SSPP 

Methodology: AP, PR, EL, JD, JY, WH, JMT 

Investigation: AP, PR, EL, JD, JY 

Visualization: AP

Project administration: GW, JMT

Supervision: GW, SSPP

Writing – original draft: AP, JY, BP, WH, JMT 

Writing – review and editing: AP, JY, BP, WH, JMT, GW, SSPP

{\bf Competing interests:}
The authors declare no competing financial or non-financial interests.

{\bf List of Supplementary Materials}

\textbf{Methods and sample fabrication:}
\begin{enumerate}
    \item[Sec.I.A:] Thin film growth. 
    \item[Sec.I.B:] Device fabrication. 
    \item[Sec.I.C:] Scanning ANE measurements.
    \item[Sec.I.D:] Near field scanning ANE measurements.
\end{enumerate} 

\textbf{Supplementary Figures:}

\begin{enumerate}
    \item [S1.] X-ray diffraction. 
    \item [S2.] Phase diagram showing the dependence of switching on the film temperature and the applied magnetic field.
    
\end{enumerate}

\newpage
\part{Supplementary materials}
\newpage
\setcounter{page}{1}   
\renewcommand\thefigure{S\arabic{figure}}    
\setcounter{figure}{0}  
\renewcommand\thesection{S\arabic{section}} 
\setcounter{section}{0}  
\renewcommand\theequation{S\arabic{equation}} 
\setcounter{equation}{0}  

\section{Methods}

\subsection{Film growth}
The $\mathrm{Mn_3Sn}$ film with a thickness of \SI{60}{nm} was deposited on a (111)-cut MgO substrate using an ultra-high vacuum magnetron sputtering system. A \SI{4}{nm} thick Ru buffer layer was deposited at \SI{500}{\degree C} at a deposition rate of \SI{0.32}{\angstrom /s}. The Mn and Sn depositions were performed at room temperature with a rate of \SI{0.68}{\angstrom /s}. After the deposition, the film was annealed at \SI{430}{\degree C}, and a \SI{3}{nm} Si protecting layer was deposited to prevent the sample from oxidation. The sample was rotated at \SI{20}{rpm} to ensure uniform growth. 

\subsection{Device fabrication}
The $\mathrm{Mn_3Sn}$ film grown on a \SI{10} {mm}$\times$\SI{10}{mm} MgO(111) substrate was cut into four parts of the size \SI{5}{mm}$\times$\SI{5}{mm} each. Two of these four pieces were utilized for the anomalous Nernst effect (ANE) based domain imaging, one for micrometer scale wide wires and the other for nanowires. For this, the wire structures were defined with lithography followed by dry etching using Ar-ion bombardment. A mask-less optical lithography system was utilized to fabricate wider wires (\SI{10}{\micro m}) that were used to image domains with the standard ANE microscope. The nanowires for domain imaging with the near-field ANE microscope were defined by an electron beam lithography system. The magneto-optical Kerr effect-based hysteresis measurement was done on the third piece of the film. The last piece was used for X-ray diffraction experiments.  

\subsection{Scanning ANE measurements}

In the standard ANE microscope setup, a laser beam with a power of \SI{1}{mW} and a wavelength of \SI{532}{nm} was focused on the sample using an objective lens. For the spatially resolved domain imaging measurements, the objective lens with a numerical aperture ($NA=0.7$) was utilized. The higher $NA$ lens gives a smaller spot size and thus better spatial resolution. For the measurements where only an average ANE signal was required (ANE-based hysteresis, and data in \textbf{Fig. S3}), the objective lens with $NA=0.4$ was used. The intensity of the laser beam was modulated using an optical chopper at a frequency of $f=\SI{600}{Hz}$. The measured ANE signal was demodulated at the same frequency by using a MFLI lock-in amplifier. The sample underneath the laser spot was scanned with a piezo stage.     

\subsection{Near field scanning ANE measurements}

For the near-field based ANE microscope, we used a scanning a near-field optical microscope (SNOM) from Neaspec. Here, a Pt tip with a resonance frequency \SI{\approx280}{KHz} was illuminated with a laser beam with a wavelength of \SI{8}{\micro m}. A parabolic mirror with $NA=0.4$ focused \SI{25}{mW} of the beam on the tip. The measured ANE signal was demodulated at the second harmonics of the tip vibration frequency using an external MFLI lock-in amplifier. 

\section{X-ray diffraction}
\begin{figure}[H]
\centering
\includegraphics[width=1\textwidth]{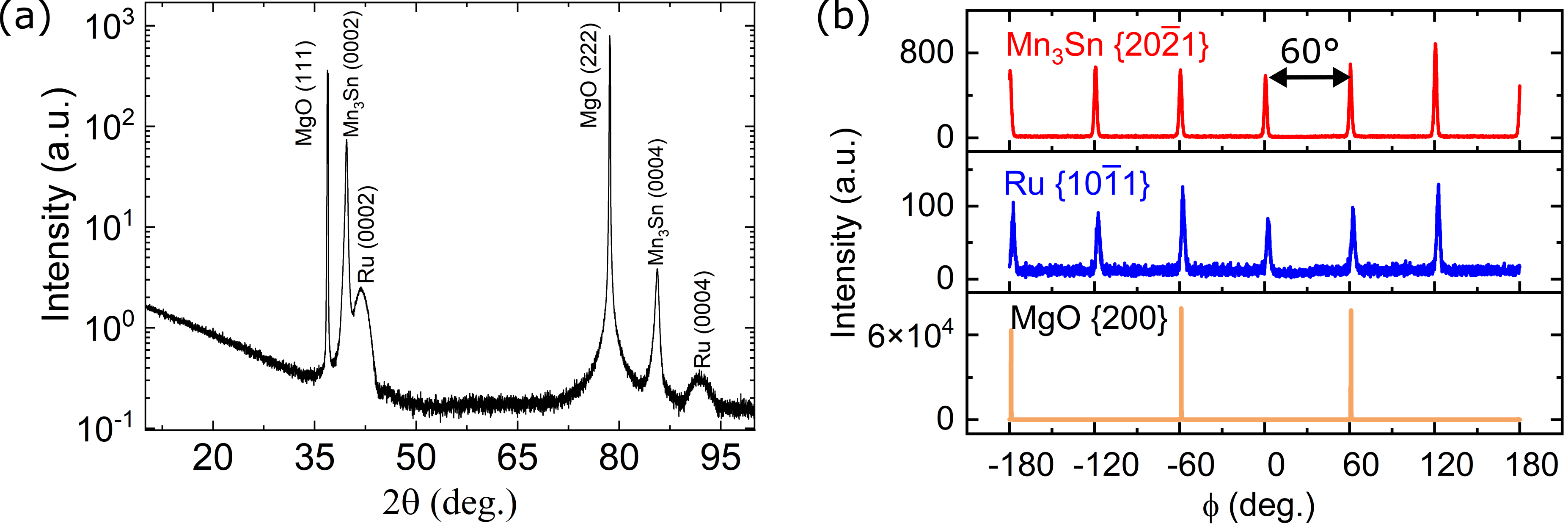}
\caption{(a) X-ray diffraction (XRD) $2\theta$ scan of the $\mathrm{Mn_3Sn}$ film. Only the (00L) family of the Bragg peaks are observed for Ru as well $\mathrm{Mn_3Sn}$, indicating c-axis-oriented epitaxial growth. (b) Azimuth XRD scans to determine the in-plane epitaxial relationship. }
\label{fig:xrd}
\end{figure}

\section{Temperature and field dependence of switching}

\begin{figure}[H]
\centering
\includegraphics[width=1\textwidth]{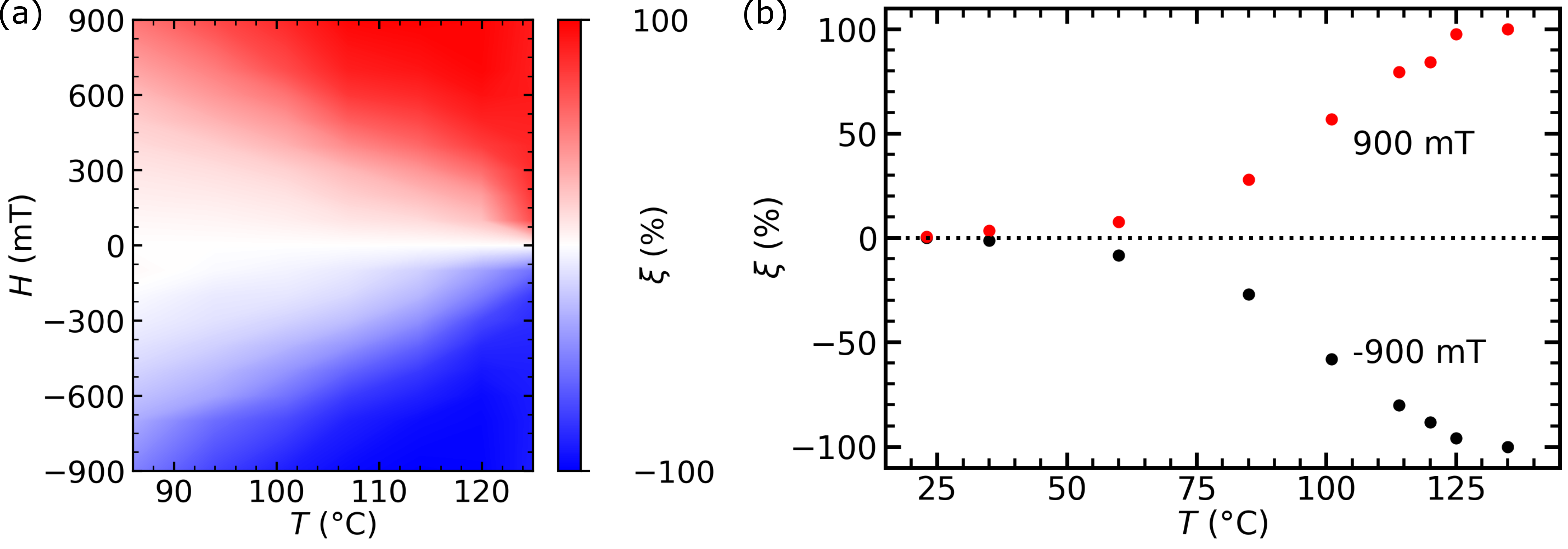}
\caption{(a) Phase diagram showing the influence of the film temperature and applied magnetic field on switching percentage ($\xi$). (b) Switching percentage for an applied field of \SI{900}{mT} (red dots) and \SI{-900}{mT} (black dots) at different temperatures.} 
\label{fig:pd_ff_fwitch}
\end{figure}

We systematically study the influence of the film temperature and applied field on the partial switching of MOOP. We define switching percentage as  $\xi(H,T)=\frac{V_{\mathrm{ANE}}(H,T)-V_{\mathrm{ANE}}(\SI{-900}{mT},\SI{125}{\degree C})}{V_{\mathrm{ANE}}(\SI{900}{mT},\SI{125}{\degree C})-V_{\mathrm{ANE}}(\SI{-900}{mT},\SI{125}{\degree C})}\times 100 $ for a positive applied field , 


The phase diagram reveals that the coercive field decreases at higher temperatures. Thus a 900 mT field can induce almost full switching above \SI{125}{\degree C} (\textbf{Fig.\ref{fig:pd_ff_fwitch}(b)}). This is because we approach the Neel temperature of Mn3Sn, making domain walls more mobile and less prone to grain boundary pinning. We demonstrate that locally heating and then field cooling is an efficient way of controlling MOOP orientation in Mn3Sn. On the other hand, the same 900 mT field cannot induce any switching at all at room temperature. Therefore, we demonstrate that the magnetic state of $\mathrm{Mn_3Sn}$ nanodevices is stable against typical environmentally-occurring stray fields. 


\begin{thebibliography}{35}
\expandafter\ifx\csname natexlab\endcsname\relax\def\natexlab#1{#1}\fi
\expandafter\ifx\csname bibnamefont\endcsname\relax
  \def\bibnamefont#1{#1}\fi
\expandafter\ifx\csname bibfnamefont\endcsname\relax
  \def\bibfnamefont#1{#1}\fi
\expandafter\ifx\csname citenamefont\endcsname\relax
  \def\citenamefont#1{#1}\fi
\expandafter\ifx\csname url\endcsname\relax
  \def\url#1{\texttt{#1}}\fi
\expandafter\ifx\csname urlprefix\endcsname\relax\def\urlprefix{URL }\fi
\providecommand{\bibinfo}[2]{#2}
\providecommand{\eprint}[2][]{\url{#2}}

\bibitem[{\citenamefont{Gomonay and Loktev}(2014)}]{AS_AIP_Gomonay2014}
\bibinfo{author}{\bibfnamefont{E.~V.} \bibnamefont{Gomonay}} \bibnamefont{and}
  \bibinfo{author}{\bibfnamefont{V.~M.} \bibnamefont{Loktev}},
  \bibinfo{journal}{Low Temperature Physics} \textbf{\bibinfo{volume}{40}},
  \bibinfo{pages}{17–35} (\bibinfo{year}{2014}), ISSN
  \bibinfo{issn}{1090-6517},
  \urlprefix\url{http://dx.doi.org/10.1063/1.4862467}.

\bibitem[{\citenamefont{MacDonald and Tsoi}(2011)}]{AS_MacDonald2011}
\bibinfo{author}{\bibfnamefont{A.~H.} \bibnamefont{MacDonald}}
  \bibnamefont{and} \bibinfo{author}{\bibfnamefont{M.}~\bibnamefont{Tsoi}},
  \bibinfo{journal}{Philosophical Transactions of the Royal Society A:
  Mathematical, Physical and Engineering Sciences}
  \textbf{\bibinfo{volume}{369}}, \bibinfo{pages}{3098–3114}
  (\bibinfo{year}{2011}), ISSN \bibinfo{issn}{1471-2962},
  \urlprefix\url{http://dx.doi.org/10.1098/rsta.2011.0014}.

\bibitem[{\citenamefont{Baltz et~al.}(2018)\citenamefont{Baltz, Manchon, Tsoi,
  Moriyama, Ono, and Tserkovnyak}}]{AS_RevModPhys_2018}
\bibinfo{author}{\bibfnamefont{V.}~\bibnamefont{Baltz}},
  \bibinfo{author}{\bibfnamefont{A.}~\bibnamefont{Manchon}},
  \bibinfo{author}{\bibfnamefont{M.}~\bibnamefont{Tsoi}},
  \bibinfo{author}{\bibfnamefont{T.}~\bibnamefont{Moriyama}},
  \bibinfo{author}{\bibfnamefont{T.}~\bibnamefont{Ono}}, \bibnamefont{and}
  \bibinfo{author}{\bibfnamefont{Y.}~\bibnamefont{Tserkovnyak}},
  \bibinfo{journal}{Rev. Mod. Phys.} \textbf{\bibinfo{volume}{90}},
  \bibinfo{pages}{015005} (\bibinfo{year}{2018}),
  \urlprefix\url{https://link.aps.org/doi/10.1103/RevModPhys.90.015005}.

\bibitem[{\citenamefont{Jungwirth et~al.}(2016)\citenamefont{Jungwirth, Marti,
  Wadley, and Wunderlich}}]{AS_Nat_NanoTech_Jungwirth2016}
\bibinfo{author}{\bibfnamefont{T.}~\bibnamefont{Jungwirth}},
  \bibinfo{author}{\bibfnamefont{X.}~\bibnamefont{Marti}},
  \bibinfo{author}{\bibfnamefont{P.}~\bibnamefont{Wadley}}, \bibnamefont{and}
  \bibinfo{author}{\bibfnamefont{J.}~\bibnamefont{Wunderlich}},
  \bibinfo{journal}{Nature Nanotechnology} \textbf{\bibinfo{volume}{11}},
  \bibinfo{pages}{231–241} (\bibinfo{year}{2016}), ISSN
  \bibinfo{issn}{1748-3395},
  \urlprefix\url{http://dx.doi.org/10.1038/nnano.2016.18}.

\bibitem[{\citenamefont{Šmejkal et~al.}(2018)\citenamefont{Šmejkal,
  Mokrousov, Yan, and MacDonald}}]{TAS_Nat_Phy_mejkal2018}
\bibinfo{author}{\bibfnamefont{L.}~\bibnamefont{Šmejkal}},
  \bibinfo{author}{\bibfnamefont{Y.}~\bibnamefont{Mokrousov}},
  \bibinfo{author}{\bibfnamefont{B.}~\bibnamefont{Yan}}, \bibnamefont{and}
  \bibinfo{author}{\bibfnamefont{A.~H.} \bibnamefont{MacDonald}},
  \bibinfo{journal}{Nature Physics} \textbf{\bibinfo{volume}{14}},
  \bibinfo{pages}{242–251} (\bibinfo{year}{2018}), ISSN
  \bibinfo{issn}{1745-2481},
  \urlprefix\url{http://dx.doi.org/10.1038/s41567-018-0064-5}.

\bibitem[{\citenamefont{Bonbien et~al.}(2021)\citenamefont{Bonbien, Zhuo,
  Salimath, Ly, Abbout, and Manchon}}]{TAS_Bonbien2021}
\bibinfo{author}{\bibfnamefont{V.}~\bibnamefont{Bonbien}},
  \bibinfo{author}{\bibfnamefont{F.}~\bibnamefont{Zhuo}},
  \bibinfo{author}{\bibfnamefont{A.}~\bibnamefont{Salimath}},
  \bibinfo{author}{\bibfnamefont{O.}~\bibnamefont{Ly}},
  \bibinfo{author}{\bibfnamefont{A.}~\bibnamefont{Abbout}}, \bibnamefont{and}
  \bibinfo{author}{\bibfnamefont{A.}~\bibnamefont{Manchon}},
  \bibinfo{journal}{Journal of Physics D: Applied Physics}
  \textbf{\bibinfo{volume}{55}}, \bibinfo{pages}{103002}
  (\bibinfo{year}{2021}), ISSN \bibinfo{issn}{1361-6463},
  \urlprefix\url{http://dx.doi.org/10.1088/1361-6463/ac28fa}.

\bibitem[{\citenamefont{Brown et~al.}(1990)\citenamefont{Brown, Nunez, Tasset,
  Forsyth, and Radhakrishna}}]{TST_Brown1990}
\bibinfo{author}{\bibfnamefont{P.~J.} \bibnamefont{Brown}},
  \bibinfo{author}{\bibfnamefont{V.}~\bibnamefont{Nunez}},
  \bibinfo{author}{\bibfnamefont{F.}~\bibnamefont{Tasset}},
  \bibinfo{author}{\bibfnamefont{J.~B.} \bibnamefont{Forsyth}},
  \bibnamefont{and}
  \bibinfo{author}{\bibfnamefont{P.}~\bibnamefont{Radhakrishna}},
  \bibinfo{journal}{Journal of Physics: Condensed Matter}
  \textbf{\bibinfo{volume}{2}}, \bibinfo{pages}{9409–9422}
  (\bibinfo{year}{1990}), ISSN \bibinfo{issn}{1361-648X},
  \urlprefix\url{http://dx.doi.org/10.1088/0953-8984/2/47/015}.

\bibitem[{\citenamefont{Park et~al.}(2018)\citenamefont{Park, Oh, Uhlířová,
  Jackson, Deák, Szunyogh, Lee, Cho, Kim, Walker et~al.}}]{TST_npj_Park2018}
\bibinfo{author}{\bibfnamefont{P.}~\bibnamefont{Park}},
  \bibinfo{author}{\bibfnamefont{J.}~\bibnamefont{Oh}},
  \bibinfo{author}{\bibfnamefont{K.}~\bibnamefont{Uhlířová}},
  \bibinfo{author}{\bibfnamefont{J.}~\bibnamefont{Jackson}},
  \bibinfo{author}{\bibfnamefont{A.}~\bibnamefont{Deák}},
  \bibinfo{author}{\bibfnamefont{L.}~\bibnamefont{Szunyogh}},
  \bibinfo{author}{\bibfnamefont{K.~H.} \bibnamefont{Lee}},
  \bibinfo{author}{\bibfnamefont{H.}~\bibnamefont{Cho}},
  \bibinfo{author}{\bibfnamefont{H.-L.} \bibnamefont{Kim}},
  \bibinfo{author}{\bibfnamefont{H.~C.} \bibnamefont{Walker}},
  \bibnamefont{et~al.}, \bibinfo{journal}{npj Quantum Materials}
  \textbf{\bibinfo{volume}{3}} (\bibinfo{year}{2018}), ISSN
  \bibinfo{issn}{2397-4648},
  \urlprefix\url{http://dx.doi.org/10.1038/s41535-018-0137-9}.

\bibitem[{\citenamefont{Kuroda et~al.}(2017)\citenamefont{Kuroda, Tomita,
  Suzuki, Bareille, Nugroho, Goswami, Ochi, Ikhlas, Nakayama, Akebi
  et~al.}}]{Weyl_ferm_nat_matKuroda2017}
\bibinfo{author}{\bibfnamefont{K.}~\bibnamefont{Kuroda}},
  \bibinfo{author}{\bibfnamefont{T.}~\bibnamefont{Tomita}},
  \bibinfo{author}{\bibfnamefont{M.-T.} \bibnamefont{Suzuki}},
  \bibinfo{author}{\bibfnamefont{C.}~\bibnamefont{Bareille}},
  \bibinfo{author}{\bibfnamefont{A.}~\bibnamefont{Nugroho}},
  \bibinfo{author}{\bibfnamefont{P.}~\bibnamefont{Goswami}},
  \bibinfo{author}{\bibfnamefont{M.}~\bibnamefont{Ochi}},
  \bibinfo{author}{\bibfnamefont{M.}~\bibnamefont{Ikhlas}},
  \bibinfo{author}{\bibfnamefont{M.}~\bibnamefont{Nakayama}},
  \bibinfo{author}{\bibfnamefont{S.}~\bibnamefont{Akebi}},
  \bibnamefont{et~al.}, \bibinfo{journal}{Nature Materials}
  \textbf{\bibinfo{volume}{16}}, \bibinfo{pages}{1090–1095}
  (\bibinfo{year}{2017}), ISSN \bibinfo{issn}{1476-4660},
  \urlprefix\url{http://dx.doi.org/10.1038/nmat4987}.

\bibitem[{\citenamefont{Šmejkal et~al.}(2017)\citenamefont{Šmejkal,
  Jungwirth, and Sinova}}]{Weyl_mejkal2017}
\bibinfo{author}{\bibfnamefont{L.}~\bibnamefont{Šmejkal}},
  \bibinfo{author}{\bibfnamefont{T.}~\bibnamefont{Jungwirth}},
  \bibnamefont{and} \bibinfo{author}{\bibfnamefont{J.}~\bibnamefont{Sinova}},
  \bibinfo{journal}{physica status solidi (RRL) – Rapid Research Letters}
  \textbf{\bibinfo{volume}{11}} (\bibinfo{year}{2017}), ISSN
  \bibinfo{issn}{1862-6270},
  \urlprefix\url{http://dx.doi.org/10.1002/pssr.201700044}.

\bibitem[{\citenamefont{K\"{u}bler and
  Felser}(2014)}]{berry_cur_field_Kbler2014}
\bibinfo{author}{\bibfnamefont{J.}~\bibnamefont{K\"{u}bler}} \bibnamefont{and}
  \bibinfo{author}{\bibfnamefont{C.}~\bibnamefont{Felser}},
  \bibinfo{journal}{EPL (Europhysics Letters)} \textbf{\bibinfo{volume}{108}},
  \bibinfo{pages}{67001} (\bibinfo{year}{2014}), ISSN
  \bibinfo{issn}{1286-4854},
  \urlprefix\url{http://dx.doi.org/10.1209/0295-5075/108/67001}.

\bibitem[{\citenamefont{Chen et~al.}(2014)\citenamefont{Chen, Niu, and
  MacDonald}}]{berry_cur_field_PRL_Chen2014}
\bibinfo{author}{\bibfnamefont{H.}~\bibnamefont{Chen}},
  \bibinfo{author}{\bibfnamefont{Q.}~\bibnamefont{Niu}}, \bibnamefont{and}
  \bibinfo{author}{\bibfnamefont{A.}~\bibnamefont{MacDonald}},
  \bibinfo{journal}{Physical Review Letters} \textbf{\bibinfo{volume}{112}}
  (\bibinfo{year}{2014}), ISSN \bibinfo{issn}{1079-7114},
  \urlprefix\url{http://dx.doi.org/10.1103/PhysRevLett.112.017205}.

\bibitem[{\citenamefont{Nakatsuji et~al.}(2015)\citenamefont{Nakatsuji,
  Kiyohara, and Higo}}]{NCAFM_AFE_Nat_Nakatsuji2015}
\bibinfo{author}{\bibfnamefont{S.}~\bibnamefont{Nakatsuji}},
  \bibinfo{author}{\bibfnamefont{N.}~\bibnamefont{Kiyohara}}, \bibnamefont{and}
  \bibinfo{author}{\bibfnamefont{T.}~\bibnamefont{Higo}},
  \bibinfo{journal}{Nature} \textbf{\bibinfo{volume}{527}},
  \bibinfo{pages}{212–215} (\bibinfo{year}{2015}), ISSN
  \bibinfo{issn}{1476-4687},
  \urlprefix\url{http://dx.doi.org/10.1038/nature15723}.

\bibitem[{\citenamefont{Ikhlas et~al.}(2017)\citenamefont{Ikhlas, Tomita,
  Koretsune, Suzuki, Nishio-Hamane, Arita, Otani, and
  Nakatsuji}}]{NCAFM_ANE_Nat_phy_Ikhlas2017}
\bibinfo{author}{\bibfnamefont{M.}~\bibnamefont{Ikhlas}},
  \bibinfo{author}{\bibfnamefont{T.}~\bibnamefont{Tomita}},
  \bibinfo{author}{\bibfnamefont{T.}~\bibnamefont{Koretsune}},
  \bibinfo{author}{\bibfnamefont{M.-T.} \bibnamefont{Suzuki}},
  \bibinfo{author}{\bibfnamefont{D.}~\bibnamefont{Nishio-Hamane}},
  \bibinfo{author}{\bibfnamefont{R.}~\bibnamefont{Arita}},
  \bibinfo{author}{\bibfnamefont{Y.}~\bibnamefont{Otani}}, \bibnamefont{and}
  \bibinfo{author}{\bibfnamefont{S.}~\bibnamefont{Nakatsuji}},
  \bibinfo{journal}{Nature Physics} \textbf{\bibinfo{volume}{13}},
  \bibinfo{pages}{1085–1090} (\bibinfo{year}{2017}), ISSN
  \bibinfo{issn}{1745-2481},
  \urlprefix\url{http://dx.doi.org/10.1038/nphys4181}.

\bibitem[{\citenamefont{Higo et~al.}(2018)\citenamefont{Higo, Man, Gopman, Wu,
  Koretsune, van~’t Erve, Kabanov, Rees, Li, Suzuki
  et~al.}}]{NCAFM_MOKE_nat_photo_Higo2018}
\bibinfo{author}{\bibfnamefont{T.}~\bibnamefont{Higo}},
  \bibinfo{author}{\bibfnamefont{H.}~\bibnamefont{Man}},
  \bibinfo{author}{\bibfnamefont{D.~B.} \bibnamefont{Gopman}},
  \bibinfo{author}{\bibfnamefont{L.}~\bibnamefont{Wu}},
  \bibinfo{author}{\bibfnamefont{T.}~\bibnamefont{Koretsune}},
  \bibinfo{author}{\bibfnamefont{O.~M.~J.} \bibnamefont{van~’t Erve}},
  \bibinfo{author}{\bibfnamefont{Y.~P.} \bibnamefont{Kabanov}},
  \bibinfo{author}{\bibfnamefont{D.}~\bibnamefont{Rees}},
  \bibinfo{author}{\bibfnamefont{Y.}~\bibnamefont{Li}},
  \bibinfo{author}{\bibfnamefont{M.-T.} \bibnamefont{Suzuki}},
  \bibnamefont{et~al.}, \bibinfo{journal}{Nature Photonics}
  \textbf{\bibinfo{volume}{12}}, \bibinfo{pages}{73–78}
  (\bibinfo{year}{2018}), ISSN \bibinfo{issn}{1749-4893},
  \urlprefix\url{http://dx.doi.org/10.1038/s41566-017-0086-z}.

\bibitem[{\citenamefont{Uchimura et~al.}(2022)\citenamefont{Uchimura, Yoon,
  Sato, Takeuchi, Kanai, Takechi, Kishi, Yamane, DuttaGupta, Ieda
  et~al.}}]{NCAFM_MOKE_Uchimura2022}
\bibinfo{author}{\bibfnamefont{T.}~\bibnamefont{Uchimura}},
  \bibinfo{author}{\bibfnamefont{J.-Y.} \bibnamefont{Yoon}},
  \bibinfo{author}{\bibfnamefont{Y.}~\bibnamefont{Sato}},
  \bibinfo{author}{\bibfnamefont{Y.}~\bibnamefont{Takeuchi}},
  \bibinfo{author}{\bibfnamefont{S.}~\bibnamefont{Kanai}},
  \bibinfo{author}{\bibfnamefont{R.}~\bibnamefont{Takechi}},
  \bibinfo{author}{\bibfnamefont{K.}~\bibnamefont{Kishi}},
  \bibinfo{author}{\bibfnamefont{Y.}~\bibnamefont{Yamane}},
  \bibinfo{author}{\bibfnamefont{S.}~\bibnamefont{DuttaGupta}},
  \bibinfo{author}{\bibfnamefont{J.}~\bibnamefont{Ieda}}, \bibnamefont{et~al.},
  \bibinfo{journal}{Applied Physics Letters} \textbf{\bibinfo{volume}{120}}
  (\bibinfo{year}{2022}), ISSN \bibinfo{issn}{1077-3118},
  \urlprefix\url{http://dx.doi.org/10.1063/5.0089355}.

\bibitem[{\citenamefont{Feng et~al.}(2015)\citenamefont{Feng, Guo, Zhou, Yao,
  and Niu}}]{NCAFM_MOKE_PhysRevB.92.144426}
\bibinfo{author}{\bibfnamefont{W.}~\bibnamefont{Feng}},
  \bibinfo{author}{\bibfnamefont{G.-Y.} \bibnamefont{Guo}},
  \bibinfo{author}{\bibfnamefont{J.}~\bibnamefont{Zhou}},
  \bibinfo{author}{\bibfnamefont{Y.}~\bibnamefont{Yao}}, \bibnamefont{and}
  \bibinfo{author}{\bibfnamefont{Q.}~\bibnamefont{Niu}},
  \bibinfo{journal}{Phys. Rev. B} \textbf{\bibinfo{volume}{92}},
  \bibinfo{pages}{144426} (\bibinfo{year}{2015}),
  \urlprefix\url{https://link.aps.org/doi/10.1103/PhysRevB.92.144426}.

\bibitem[{\citenamefont{Sakamoto et~al.}(2021)\citenamefont{Sakamoto, Higo,
  Shiga, Amemiya, Nakatsuji, and Miwa}}]{NCAFM_XMCD_PhysRevB.104.134431}
\bibinfo{author}{\bibfnamefont{S.}~\bibnamefont{Sakamoto}},
  \bibinfo{author}{\bibfnamefont{T.}~\bibnamefont{Higo}},
  \bibinfo{author}{\bibfnamefont{M.}~\bibnamefont{Shiga}},
  \bibinfo{author}{\bibfnamefont{K.}~\bibnamefont{Amemiya}},
  \bibinfo{author}{\bibfnamefont{S.}~\bibnamefont{Nakatsuji}},
  \bibnamefont{and} \bibinfo{author}{\bibfnamefont{S.}~\bibnamefont{Miwa}},
  \bibinfo{journal}{Phys. Rev. B} \textbf{\bibinfo{volume}{104}},
  \bibinfo{pages}{134431} (\bibinfo{year}{2021}),
  \urlprefix\url{https://link.aps.org/doi/10.1103/PhysRevB.104.134431}.

\bibitem[{\citenamefont{Kimata et~al.}(2021)\citenamefont{Kimata, Sasabe,
  Kurita, Yamasaki, Tabata, Yokoyama, Kotani, Ikhlas, Tomita, Amemiya
  et~al.}}]{NCAFM_XMCD_Nat_com_Kimata2021}
\bibinfo{author}{\bibfnamefont{M.}~\bibnamefont{Kimata}},
  \bibinfo{author}{\bibfnamefont{N.}~\bibnamefont{Sasabe}},
  \bibinfo{author}{\bibfnamefont{K.}~\bibnamefont{Kurita}},
  \bibinfo{author}{\bibfnamefont{Y.}~\bibnamefont{Yamasaki}},
  \bibinfo{author}{\bibfnamefont{C.}~\bibnamefont{Tabata}},
  \bibinfo{author}{\bibfnamefont{Y.}~\bibnamefont{Yokoyama}},
  \bibinfo{author}{\bibfnamefont{Y.}~\bibnamefont{Kotani}},
  \bibinfo{author}{\bibfnamefont{M.}~\bibnamefont{Ikhlas}},
  \bibinfo{author}{\bibfnamefont{T.}~\bibnamefont{Tomita}},
  \bibinfo{author}{\bibfnamefont{K.}~\bibnamefont{Amemiya}},
  \bibnamefont{et~al.}, \bibinfo{journal}{Nature Communications}
  \textbf{\bibinfo{volume}{12}} (\bibinfo{year}{2021}), ISSN
  \bibinfo{issn}{2041-1723},
  \urlprefix\url{http://dx.doi.org/10.1038/s41467-021-25834-7}.

\bibitem[{\citenamefont{Chen et~al.}(2023)\citenamefont{Chen, Higo, Tanaka,
  Nomoto, Tsai, Idzuchi, Shiga, Sakamoto, Ando, Kosaki et~al.}}]{MOOP_Chen2023}
\bibinfo{author}{\bibfnamefont{X.}~\bibnamefont{Chen}},
  \bibinfo{author}{\bibfnamefont{T.}~\bibnamefont{Higo}},
  \bibinfo{author}{\bibfnamefont{K.}~\bibnamefont{Tanaka}},
  \bibinfo{author}{\bibfnamefont{T.}~\bibnamefont{Nomoto}},
  \bibinfo{author}{\bibfnamefont{H.}~\bibnamefont{Tsai}},
  \bibinfo{author}{\bibfnamefont{H.}~\bibnamefont{Idzuchi}},
  \bibinfo{author}{\bibfnamefont{M.}~\bibnamefont{Shiga}},
  \bibinfo{author}{\bibfnamefont{S.}~\bibnamefont{Sakamoto}},
  \bibinfo{author}{\bibfnamefont{R.}~\bibnamefont{Ando}},
  \bibinfo{author}{\bibfnamefont{H.}~\bibnamefont{Kosaki}},
  \bibnamefont{et~al.}, \bibinfo{journal}{Nature}
  \textbf{\bibinfo{volume}{613}}, \bibinfo{pages}{490–495}
  (\bibinfo{year}{2023}), ISSN \bibinfo{issn}{1476-4687},
  \urlprefix\url{http://dx.doi.org/10.1038/s41586-022-05463-w}.

\bibitem[{\citenamefont{Nakatsuji and Arita}(2022)}]{MOOP_Nakatsuji2022}
\bibinfo{author}{\bibfnamefont{S.}~\bibnamefont{Nakatsuji}} \bibnamefont{and}
  \bibinfo{author}{\bibfnamefont{R.}~\bibnamefont{Arita}},
  \bibinfo{journal}{Annual Review of Condensed Matter Physics}
  \textbf{\bibinfo{volume}{13}}, \bibinfo{pages}{119–142}
  (\bibinfo{year}{2022}), ISSN \bibinfo{issn}{1947-5462},
  \urlprefix\url{http://dx.doi.org/10.1146/annurev-conmatphys-031620-103859}.

\bibitem[{\citenamefont{Chen et~al.}(2021)\citenamefont{Chen, Tomita, Minami,
  Fu, Koretsune, Kitatani, Muhammad, Nishio-Hamane, Ishii, Ishii
  et~al.}}]{`moop_resp_Chen2021}
\bibinfo{author}{\bibfnamefont{T.}~\bibnamefont{Chen}},
  \bibinfo{author}{\bibfnamefont{T.}~\bibnamefont{Tomita}},
  \bibinfo{author}{\bibfnamefont{S.}~\bibnamefont{Minami}},
  \bibinfo{author}{\bibfnamefont{M.}~\bibnamefont{Fu}},
  \bibinfo{author}{\bibfnamefont{T.}~\bibnamefont{Koretsune}},
  \bibinfo{author}{\bibfnamefont{M.}~\bibnamefont{Kitatani}},
  \bibinfo{author}{\bibfnamefont{I.}~\bibnamefont{Muhammad}},
  \bibinfo{author}{\bibfnamefont{D.}~\bibnamefont{Nishio-Hamane}},
  \bibinfo{author}{\bibfnamefont{R.}~\bibnamefont{Ishii}},
  \bibinfo{author}{\bibfnamefont{F.}~\bibnamefont{Ishii}},
  \bibnamefont{et~al.}, \bibinfo{journal}{Nature Communications}
  \textbf{\bibinfo{volume}{12}} (\bibinfo{year}{2021}), ISSN
  \bibinfo{issn}{2041-1723},
  \urlprefix\url{http://dx.doi.org/10.1038/s41467-020-20838-1}.

\bibitem[{\citenamefont{Cheong et~al.}(2020)\citenamefont{Cheong, Fiebig, Wu,
  Chapon, and Kiryukhin}}]{dome_size_Cheong2020}
\bibinfo{author}{\bibfnamefont{S.-W.} \bibnamefont{Cheong}},
  \bibinfo{author}{\bibfnamefont{M.}~\bibnamefont{Fiebig}},
  \bibinfo{author}{\bibfnamefont{W.}~\bibnamefont{Wu}},
  \bibinfo{author}{\bibfnamefont{L.}~\bibnamefont{Chapon}}, \bibnamefont{and}
  \bibinfo{author}{\bibfnamefont{V.}~\bibnamefont{Kiryukhin}},
  \bibinfo{journal}{npj Quantum Materials} \textbf{\bibinfo{volume}{5}}
  (\bibinfo{year}{2020}), ISSN \bibinfo{issn}{2397-4648},
  \urlprefix\url{http://dx.doi.org/10.1038/s41535-019-0204-x}.

\bibitem[{\citenamefont{Krizek et~al.}(2022)\citenamefont{Krizek, Reimers,
  Kašpar, Marmodoro, Michalička, Man, Edstr\"{o}m, Amin, Edmonds, Campion
  et~al.}}]{dom_size_Krizek2022}
\bibinfo{author}{\bibfnamefont{F.}~\bibnamefont{Krizek}},
  \bibinfo{author}{\bibfnamefont{S.}~\bibnamefont{Reimers}},
  \bibinfo{author}{\bibfnamefont{Z.}~\bibnamefont{Kašpar}},
  \bibinfo{author}{\bibfnamefont{A.}~\bibnamefont{Marmodoro}},
  \bibinfo{author}{\bibfnamefont{J.}~\bibnamefont{Michalička}},
  \bibinfo{author}{\bibfnamefont{O.}~\bibnamefont{Man}},
  \bibinfo{author}{\bibfnamefont{A.}~\bibnamefont{Edstr\"{o}m}},
  \bibinfo{author}{\bibfnamefont{O.~J.} \bibnamefont{Amin}},
  \bibinfo{author}{\bibfnamefont{K.~W.} \bibnamefont{Edmonds}},
  \bibinfo{author}{\bibfnamefont{R.~P.} \bibnamefont{Campion}},
  \bibnamefont{et~al.}, \bibinfo{journal}{Science Advances}
  \textbf{\bibinfo{volume}{8}} (\bibinfo{year}{2022}), ISSN
  \bibinfo{issn}{2375-2548},
  \urlprefix\url{http://dx.doi.org/10.1126/sciadv.abn3535}.

\bibitem[{\citenamefont{Reichlova et~al.}(2019)\citenamefont{Reichlova, Janda,
  Godinho, Markou, Kriegner, Schlitz, Zelezny, Soban, Bejarano, Schultheiss
  et~al.}}]{SANE_Reichlova2019}
\bibinfo{author}{\bibfnamefont{H.}~\bibnamefont{Reichlova}},
  \bibinfo{author}{\bibfnamefont{T.}~\bibnamefont{Janda}},
  \bibinfo{author}{\bibfnamefont{J.}~\bibnamefont{Godinho}},
  \bibinfo{author}{\bibfnamefont{A.}~\bibnamefont{Markou}},
  \bibinfo{author}{\bibfnamefont{D.}~\bibnamefont{Kriegner}},
  \bibinfo{author}{\bibfnamefont{R.}~\bibnamefont{Schlitz}},
  \bibinfo{author}{\bibfnamefont{J.}~\bibnamefont{Zelezny}},
  \bibinfo{author}{\bibfnamefont{Z.}~\bibnamefont{Soban}},
  \bibinfo{author}{\bibfnamefont{M.}~\bibnamefont{Bejarano}},
  \bibinfo{author}{\bibfnamefont{H.}~\bibnamefont{Schultheiss}},
  \bibnamefont{et~al.}, \bibinfo{journal}{Nature Communications}
  \textbf{\bibinfo{volume}{10}} (\bibinfo{year}{2019}), ISSN
  \bibinfo{issn}{2041-1723},
  \urlprefix\url{http://dx.doi.org/10.1038/s41467-019-13391-z}.

\bibitem[{\citenamefont{Johnson et~al.}(2022)\citenamefont{Johnson, Kimák,
  Zemen, Šobáň, Schmoranzerová, Godinho, Němec, Beckert, Reichlová,
  Boldrin et~al.}}]{SANE_Johnson2022}
\bibinfo{author}{\bibfnamefont{F.}~\bibnamefont{Johnson}},
  \bibinfo{author}{\bibfnamefont{J.}~\bibnamefont{Kimák}},
  \bibinfo{author}{\bibfnamefont{J.}~\bibnamefont{Zemen}},
  \bibinfo{author}{\bibfnamefont{Z.}~\bibnamefont{Šobáň}},
  \bibinfo{author}{\bibfnamefont{E.}~\bibnamefont{Schmoranzerová}},
  \bibinfo{author}{\bibfnamefont{J.}~\bibnamefont{Godinho}},
  \bibinfo{author}{\bibfnamefont{P.}~\bibnamefont{Němec}},
  \bibinfo{author}{\bibfnamefont{S.}~\bibnamefont{Beckert}},
  \bibinfo{author}{\bibfnamefont{H.}~\bibnamefont{Reichlová}},
  \bibinfo{author}{\bibfnamefont{D.}~\bibnamefont{Boldrin}},
  \bibnamefont{et~al.}, \bibinfo{journal}{Applied Physics Letters}
  \textbf{\bibinfo{volume}{120}} (\bibinfo{year}{2022}), ISSN
  \bibinfo{issn}{1077-3118},
  \urlprefix\url{http://dx.doi.org/10.1063/5.0091257}.

\bibitem[{\citenamefont{Isshiki et~al.}(2024)\citenamefont{Isshiki, Budai,
  Kobayashi, Uesugi, Higo, Nakatsuji, and
  Otani}}]{SNOM_otani_PhysRevLett.132.216702}
\bibinfo{author}{\bibfnamefont{H.}~\bibnamefont{Isshiki}},
  \bibinfo{author}{\bibfnamefont{N.}~\bibnamefont{Budai}},
  \bibinfo{author}{\bibfnamefont{A.}~\bibnamefont{Kobayashi}},
  \bibinfo{author}{\bibfnamefont{R.}~\bibnamefont{Uesugi}},
  \bibinfo{author}{\bibfnamefont{T.}~\bibnamefont{Higo}},
  \bibinfo{author}{\bibfnamefont{S.}~\bibnamefont{Nakatsuji}},
  \bibnamefont{and} \bibinfo{author}{\bibfnamefont{Y.}~\bibnamefont{Otani}},
  \bibinfo{journal}{Phys. Rev. Lett.} \textbf{\bibinfo{volume}{132}},
  \bibinfo{pages}{216702} (\bibinfo{year}{2024}),
  \urlprefix\url{https://link.aps.org/doi/10.1103/PhysRevLett.132.216702}.

\bibitem[{\citenamefont{Pal et~al.}(2022)\citenamefont{Pal, Hazra, G\"{o}bel,
  Jeon, Pandeya, Chakraborty, Busch, Srivastava, Deniz, Taylor
  et~al.}}]{six_dom_Pal2022}
\bibinfo{author}{\bibfnamefont{B.}~\bibnamefont{Pal}},
  \bibinfo{author}{\bibfnamefont{B.~K.} \bibnamefont{Hazra}},
  \bibinfo{author}{\bibfnamefont{B.}~\bibnamefont{G\"{o}bel}},
  \bibinfo{author}{\bibfnamefont{J.-C.} \bibnamefont{Jeon}},
  \bibinfo{author}{\bibfnamefont{A.~K.} \bibnamefont{Pandeya}},
  \bibinfo{author}{\bibfnamefont{A.}~\bibnamefont{Chakraborty}},
  \bibinfo{author}{\bibfnamefont{O.}~\bibnamefont{Busch}},
  \bibinfo{author}{\bibfnamefont{A.~K.} \bibnamefont{Srivastava}},
  \bibinfo{author}{\bibfnamefont{H.}~\bibnamefont{Deniz}},
  \bibinfo{author}{\bibfnamefont{J.~M.} \bibnamefont{Taylor}},
  \bibnamefont{et~al.}, \bibinfo{journal}{Science Advances}
  \textbf{\bibinfo{volume}{8}} (\bibinfo{year}{2022}), ISSN
  \bibinfo{issn}{2375-2548},
  \urlprefix\url{http://dx.doi.org/10.1126/sciadv.abo5930}.

\bibitem[{\citenamefont{Sugimoto et~al.}(2020)\citenamefont{Sugimoto, Nakatani,
  Yamane, Ikhlas, Kondou, Kimata, Tomita, Nakatsuji, and
  Otani}}]{six_Sugimoto2020}
\bibinfo{author}{\bibfnamefont{S.}~\bibnamefont{Sugimoto}},
  \bibinfo{author}{\bibfnamefont{Y.}~\bibnamefont{Nakatani}},
  \bibinfo{author}{\bibfnamefont{Y.}~\bibnamefont{Yamane}},
  \bibinfo{author}{\bibfnamefont{M.}~\bibnamefont{Ikhlas}},
  \bibinfo{author}{\bibfnamefont{K.}~\bibnamefont{Kondou}},
  \bibinfo{author}{\bibfnamefont{M.}~\bibnamefont{Kimata}},
  \bibinfo{author}{\bibfnamefont{T.}~\bibnamefont{Tomita}},
  \bibinfo{author}{\bibfnamefont{S.}~\bibnamefont{Nakatsuji}},
  \bibnamefont{and} \bibinfo{author}{\bibfnamefont{Y.}~\bibnamefont{Otani}},
  \bibinfo{journal}{Communications Physics} \textbf{\bibinfo{volume}{3}}
  (\bibinfo{year}{2020}), ISSN \bibinfo{issn}{2399-3650},
  \urlprefix\url{http://dx.doi.org/10.1038/s42005-020-0381-8}.

\bibitem[{\citenamefont{Pandey et~al.}(2024)\citenamefont{Pandey, Deka, Yoon,
  Koerner, Dreyer, Taylor, Parkin, and Woltersdorf}}]{arxiv}
\bibinfo{author}{\bibfnamefont{A.}~\bibnamefont{Pandey}},
  \bibinfo{author}{\bibfnamefont{J.}~\bibnamefont{Deka}},
  \bibinfo{author}{\bibfnamefont{J.}~\bibnamefont{Yoon}},
  \bibinfo{author}{\bibfnamefont{C.}~\bibnamefont{Koerner}},
  \bibinfo{author}{\bibfnamefont{R.}~\bibnamefont{Dreyer}},
  \bibinfo{author}{\bibfnamefont{J.~M.} \bibnamefont{Taylor}},
  \bibinfo{author}{\bibfnamefont{S.~S.~P.} \bibnamefont{Parkin}},
  \bibnamefont{and}
  \bibinfo{author}{\bibfnamefont{G.}~\bibnamefont{Woltersdorf}},
  \emph{\bibinfo{title}{Anomalous nernst effect based near field imaging of
  magnetic nanostructures}} (\bibinfo{year}{2024}),
  \urlprefix\url{https://arxiv.org/abs/2407.13028}.

\bibitem[{\citenamefont{Grzybowski et~al.}(2017)\citenamefont{Grzybowski,
  Wadley, Edmonds, Beardsley, Hills, Campion, Gallagher, Chauhan, Novak,
  Jungwirth et~al.}}]{PEEM_SOT_CuMnAs_PhysRevLett.118.057701}
\bibinfo{author}{\bibfnamefont{M.~J.} \bibnamefont{Grzybowski}},
  \bibinfo{author}{\bibfnamefont{P.}~\bibnamefont{Wadley}},
  \bibinfo{author}{\bibfnamefont{K.~W.} \bibnamefont{Edmonds}},
  \bibinfo{author}{\bibfnamefont{R.}~\bibnamefont{Beardsley}},
  \bibinfo{author}{\bibfnamefont{V.}~\bibnamefont{Hills}},
  \bibinfo{author}{\bibfnamefont{R.~P.} \bibnamefont{Campion}},
  \bibinfo{author}{\bibfnamefont{B.~L.} \bibnamefont{Gallagher}},
  \bibinfo{author}{\bibfnamefont{J.~S.} \bibnamefont{Chauhan}},
  \bibinfo{author}{\bibfnamefont{V.}~\bibnamefont{Novak}},
  \bibinfo{author}{\bibfnamefont{T.}~\bibnamefont{Jungwirth}},
  \bibnamefont{et~al.}, \bibinfo{journal}{Phys. Rev. Lett.}
  \textbf{\bibinfo{volume}{118}}, \bibinfo{pages}{057701}
  (\bibinfo{year}{2017}),
  \urlprefix\url{https://link.aps.org/doi/10.1103/PhysRevLett.118.057701}.

\bibitem[{\citenamefont{Bodnar et~al.}(2019)\citenamefont{Bodnar, Filianina,
  Bommanaboyena, Forrest, Maccherozzi, Sapozhnik, Skourski, Kl\"aui, and
  Jourdan}}]{PEEM_SOT_Mn2AuPhysRevB.99.140409}
\bibinfo{author}{\bibfnamefont{S.~Y.} \bibnamefont{Bodnar}},
  \bibinfo{author}{\bibfnamefont{M.}~\bibnamefont{Filianina}},
  \bibinfo{author}{\bibfnamefont{S.~P.} \bibnamefont{Bommanaboyena}},
  \bibinfo{author}{\bibfnamefont{T.}~\bibnamefont{Forrest}},
  \bibinfo{author}{\bibfnamefont{F.}~\bibnamefont{Maccherozzi}},
  \bibinfo{author}{\bibfnamefont{A.~A.} \bibnamefont{Sapozhnik}},
  \bibinfo{author}{\bibfnamefont{Y.}~\bibnamefont{Skourski}},
  \bibinfo{author}{\bibfnamefont{M.}~\bibnamefont{Kl\"aui}}, \bibnamefont{and}
  \bibinfo{author}{\bibfnamefont{M.}~\bibnamefont{Jourdan}},
  \bibinfo{journal}{Phys. Rev. B} \textbf{\bibinfo{volume}{99}},
  \bibinfo{pages}{140409} (\bibinfo{year}{2019}),
  \urlprefix\url{https://link.aps.org/doi/10.1103/PhysRevB.99.140409}.

\bibitem[{\citenamefont{Sapozhnik et~al.}(2018)\citenamefont{Sapozhnik,
  Filianina, Bodnar, Lamirand, Mawass, Skourski, Elmers, Zabel, Kl\"aui, and
  Jourdan}}]{PEEM_field_Mn2AuPhysRevB.97.134429}
\bibinfo{author}{\bibfnamefont{A.~A.} \bibnamefont{Sapozhnik}},
  \bibinfo{author}{\bibfnamefont{M.}~\bibnamefont{Filianina}},
  \bibinfo{author}{\bibfnamefont{S.~Y.} \bibnamefont{Bodnar}},
  \bibinfo{author}{\bibfnamefont{A.}~\bibnamefont{Lamirand}},
  \bibinfo{author}{\bibfnamefont{M.-A.} \bibnamefont{Mawass}},
  \bibinfo{author}{\bibfnamefont{Y.}~\bibnamefont{Skourski}},
  \bibinfo{author}{\bibfnamefont{H.-J.} \bibnamefont{Elmers}},
  \bibinfo{author}{\bibfnamefont{H.}~\bibnamefont{Zabel}},
  \bibinfo{author}{\bibfnamefont{M.}~\bibnamefont{Kl\"aui}}, \bibnamefont{and}
  \bibinfo{author}{\bibfnamefont{M.}~\bibnamefont{Jourdan}},
  \bibinfo{journal}{Phys. Rev. B} \textbf{\bibinfo{volume}{97}},
  \bibinfo{pages}{134429} (\bibinfo{year}{2018}),
  \urlprefix\url{https://link.aps.org/doi/10.1103/PhysRevB.97.134429}.

\bibitem[{\citenamefont{Pfitzner et~al.}(2018)\citenamefont{Pfitzner, Hu,
  Schumacher, Hoehl, Venkateshvaran, Cubukcu, Liao, Auffret, Heberle,
  Wunderlich et~al.}}]{ANE_PMA_edge_AIP_adv_2018}
\bibinfo{author}{\bibfnamefont{E.}~\bibnamefont{Pfitzner}},
  \bibinfo{author}{\bibfnamefont{X.}~\bibnamefont{Hu}},
  \bibinfo{author}{\bibfnamefont{H.~W.} \bibnamefont{Schumacher}},
  \bibinfo{author}{\bibfnamefont{A.}~\bibnamefont{Hoehl}},
  \bibinfo{author}{\bibfnamefont{D.}~\bibnamefont{Venkateshvaran}},
  \bibinfo{author}{\bibfnamefont{M.}~\bibnamefont{Cubukcu}},
  \bibinfo{author}{\bibfnamefont{J.-W.} \bibnamefont{Liao}},
  \bibinfo{author}{\bibfnamefont{S.}~\bibnamefont{Auffret}},
  \bibinfo{author}{\bibfnamefont{J.}~\bibnamefont{Heberle}},
  \bibinfo{author}{\bibfnamefont{J.}~\bibnamefont{Wunderlich}},
  \bibnamefont{et~al.}, \bibinfo{journal}{AIP Advances}
  \textbf{\bibinfo{volume}{8}}, \bibinfo{pages}{125329} (\bibinfo{year}{2018}),
  ISSN \bibinfo{issn}{2158-3226},
  \urlprefix\url{https://doi.org/10.1063/1.5054382}.

\bibitem[{\citenamefont{Filippou et~al.}(2018)\citenamefont{Filippou, Jeong,
  Ferrante, Yang, Topuria, Samant, and Parkin}}]{Templated_growth_Filippou2018}
\bibinfo{author}{\bibfnamefont{P.~C.} \bibnamefont{Filippou}},
  \bibinfo{author}{\bibfnamefont{J.}~\bibnamefont{Jeong}},
  \bibinfo{author}{\bibfnamefont{Y.}~\bibnamefont{Ferrante}},
  \bibinfo{author}{\bibfnamefont{S.-H.} \bibnamefont{Yang}},
  \bibinfo{author}{\bibfnamefont{T.}~\bibnamefont{Topuria}},
  \bibinfo{author}{\bibfnamefont{M.~G.} \bibnamefont{Samant}},
  \bibnamefont{and} \bibinfo{author}{\bibfnamefont{S.~S.~P.}
  \bibnamefont{Parkin}}, \bibinfo{journal}{Nature Communications}
  \textbf{\bibinfo{volume}{9}} (\bibinfo{year}{2018}), ISSN
  \bibinfo{issn}{2041-1723},
  \urlprefix\url{http://dx.doi.org/10.1038/s41467-018-07091-3}.

\end{thebibliography}
\end{document}